\documentstyle [12pt]{article}
\begin{document}

\hfill {WM-95-104}

\hfill {TIFR/TH/95-20}

\hfill {June 1995}

\vskip 1in   \baselineskip 24pt
{
\Large

   \bigskip
   \centerline{New Bounds on R-parity Violating Couplings} }
 \vskip .8in

\centerline{Carl E. Carlson${}^1$, Probir Roy${}^2$ and Marc
Sher${}^1$ }
\bigskip
\centerline {${}^1$\it Physics Department, College of William and
Mary, Williamsburg, VA 23187, USA}
\centerline {${}^2$\it Tata Institute of Fundamental Research,
Homi Bhabha Rd., Bombay 400 005, INDIA}
\vskip 1in

{\narrower\narrower  We use information from rare nonleptonic
decays of heavy-quark mesons to put new bounds on the magnitudes of
certain product combinations of baryon nonconserving R-parity
violating couplings in supersymmetric models.  Product combinations
of lepton {\it and} baryon nonconserving R-parity violating
couplings are also considered in the light of existing bounds on
nucleon decay.  Contrary to popular impression, a few such
combinations are shown to remain essentially unconstrained.}

\newpage

\def\beq{\begin{equation}}
\def\eeq{\end{equation}}
\def\lm{\lambda}
\def\lusd{\lm_{usd}''}
\def\lubd{\lm_{ubd}''}
\def\lubs{\lm_{ubs}''}
\def\lcsd{\lm_{csd}''}
\def\lcbd{\lm_{cbd}''}
\def\lcbs{\lm_{cbs}''}
\def\ltsd{\lm_{tsd}''}
\def\ltbd{\lm_{tbd}''}
\def\ltbs{\lm_{tbs}''}

Though the minimal supersymmetric standard model (MSSM)~\cite{one}
is a leading candidate for new physics beyond the standard model,
the conservation of R-parity, $R_p$, which is assumed in the model
has no real theoretical justification.  This has motivated many
authors~\cite{two} to consider alternatives in which $R_p$ is
explicitly broken.  In such models, sparticles can decay into
non-supersymmetric particles alone, leading to novel signatures in
search experiments and unusual decay processes.

The most general $R_p$-violating superpotential that one can write
with the MSSM superfields, in the usual notation, is
\beq
W=\lm_{ijk}L_iL_j\bar{E}_k+\lm_{ijk}'L_iQ_j\bar{D}_k
+\lm_{ijk}''\bar{U}_i\bar{D}_j\bar{D}_k
\eeq
Here, $i,j,k$ are generation indices and we have rotated away a
term of the form $\mu_{ij}L_iH_j$.  Since the $\lm_{ijk}$ term
is symmetric under exchange of $i$ and $j$, and antisymmetric in
color, it must be antisymmetric in flavor, thus we have
$\lm_{ijk}=-\lm_{jik}$.  Similarly, $\lm_{ijk}''=-\lm_{ikj}$.  The
number of couplings is then 36 lepton nonconserving couplings (9
of the $\lm$ type and 27 of the $\lm'$ type) and 9 baryon
nonconserving couplings (all of the $\lm''$ type) in total.

It is generally thought that $\lm$, $\lm'$ type couplings cannot
coexist with $\lm''$ type couplings since both baryon and lepton
number violations would lead to too rapid a proton decay.  For this
reason, previous authors have considered either lepton
nonconserving {\it or} baryon nonconserving couplings, but not
both.  We will first make this assumption and consider the baryon
number violating $\lm''$ couplings alone, since most of the
earlier effort has been focused on $\lm$ and $\lm'$
couplings~\cite{three}.  Later, we will examine proton decay in the
presence of all three types of couplings.
Severe constraints on $R_p$-violating couplings can be obtained
by requiring that the cosmological baryon asymmetry not be
washed out\cite{bary}, but it is possible to
evade these bounds~\cite{dreiner}.

Our philosophy throughout will be that we expect the couplings
involving third generation fields to be the largest.  There are
two reasons for this.  The only other Yukawa couplings in the
model (the Yukawa couplings of the quarks to the Higgs fields)
exhibit a very strongly hierarchical generational structure and
thus one would expect the R-parity violating couplings to do so as
well.  Second, many models result in the scalar top being the
lightest of the squarks, and thus processes involving the scalar top
may not be as suppressed by large propagators.  We will thus
concentrate on couplings involving higher generations, but will
keep our bounds as general as possible.

We can write the 9 different $\lm''$ couplings as $\ltbs, \ltbd,
\ltsd, \lcbs, \lcbd, \lcsd, \lubs, \lubd$ and $\lusd$.  Let us
first recount the existing constraints on these couplings.
 Brahmachari and Roy~\cite{br} showed that the requirement of
perturbative unification typically places a bound of between
$1.10$ and
$1.25$ on many of the couplings.  This was generalized to all the
couplings by Goity and Sher~\cite{goity}.  The latter also showed,
following earlier work~\cite{zw, barbieri},   that
$|\lusd|$  can be strongly bounded by the nonobservation of
double nucleon decay into two kaons (such as
${}^{16}O\rightarrow {}^{14}C\ K^+ K^+$, which would have been seen
in water Cerenkov detectors), and  $|\lubd|$  can be
strongly bounded by the nonobservation of neutron-antineutron
oscillations. Their bounds, for squark masses of 300 GeV, were
$|\lubd| < 5\times 10^{-3}$ and $|\lusd| < 10^{-6}$.  In the
work of Barbieri and Masiero~\cite{barbieri}, some bounds on
products of couplings were obtained by considering $K$-$\bar{K}$
mixing; these bounds will be discussed shortly.  Finally, bounds
from the $\bar{b}b$ induced vertex correction to the decay
of the $Z$ into two charged leptons
 have recently been obtained~\cite{cern}; though
potentially interesting, with present data they are not
significantly better than the bound from perturbative unification.

In this Letter, we note that many additional and interesting
bounds on the $\lm''$ couplings can be obtained by considering rare
two-body nonleptonic decays of heavy-quark mesons.  We shall
mostly consider B decays, but also, in some
cases, D decays.  Let us begin by considering the implications for
$\lambda''$ couplings from such processes..  Since
lepton number is assumed to be conserved, only $\Delta B=0$ and
$\Delta B=2$ decays can occur. Thus any bounds will be on the
 products of two $\lambda''$ couplings.  Furthermore,
since any B-decay will change the number of ``b-flavors'' by one
unit, bounds from there  will apply to products of the form
$\lambda_{{u_i}bs}\lambda_{{u_j}sd}$ or
$\lambda_{{u_i}bd}\lambda_{{u_j}sd}$. We first consider
$\Delta B=0$ (baryon number conserving) decays which actually give
the best bounds and later comment on the $\Delta B=2$ processes.

In our calculation of two-body nonleptonic decays of heavy-quark
mesons, we follow the computational method of Carlson and
Milana~\cite{cm} which is based upon the formalism of Brodsky and
Lepage~\cite{bl}.  First, we neglect all light meson masses
($\pi$'s, K's).  Then we make use of the fact that the relative
momentum between the quark and the antiquark for each of the
$q\bar{q}$ pairs within the
decaying meson and within each of the final state mesons is low.
 The large quark momentum transfers, needed to
move a quark from one meson to another or to produce a
$q\bar{q}$ pair which enter different mesons, can be caused
either by single-gluon
exchange or through the emission of a virtual squark.  However,
both are needed in each diagram for the correct distribution of
relative momentum, so we consider diagrams that involve both.
Alternatively, one could consider the full meson wave
functions~\cite{isg} with the tail of the wave functions being
crucial; if properly done, this should be equivalent to the
previous calculation since at high enough relative momentum the
tail is indeed given by 1-gluon exchange.

First, take the decay $B^+\rightarrow \bar{K}^oK^+$ (or
equivalently, $B^-\rightarrow K^oK^-$).  This has an extremely
small rate in the Standard Model, being penguin-suppressed and
also reduced by the small CKM element $V_{ub}$ in the amplitude.
The dominant diagrams contributing to this process for nonzero
$\lm''$ couplings are shown in Fig. 1.  In each of these diagrams,
the gluon is spacelike and the squark (of charge $2/3$) is
timelike.  This generates from the gluon propagator an overall
enhancement factor of $m_B/(m_B-m_b)\simeq 10$ in the amplitude,
$m_B$ and $m_b$ being the B-meson and b-quark masses
respectively.  (This factor is just the inverse of the fraction of
the B-momentum assigned to the light quark.)  One can draw similar
diagrams interchanging the squark and gluon internal lines and
appropriately relabeling the quark lines.  Each of these latter
diagrams would have a timelike gluon and a spacelike squark (now
of charge $-1/3$). For these, however, the overall factor of
$m_B/(m_B-m_b)$ does not materialize so that contributions from
these diagrams are significantly subdominant and can be
neglected.  It is of interest to point out that the decay
$B^o\rightarrow K^+K^-$ can proceed only through these latter
diagrams, with the squark having charge $2/3$, and therefore
yields a significantly weaker bound on the same product of $\lm''$
couplings than does $B^+\rightarrow \bar{K}^oK^+$, in spite
of stronger experimental limits on the branching ratio.

Our result is most conveniently expressed as a ratio of
$\Gamma(B^+\rightarrow \bar{K}^oK^+)$ to the partial width of
another $B^+$ decay channel (specifically $B^+\rightarrow K^+
J/\psi$) that proceeds unsuppressed in the Standard Model.  This
description eliminates many of the uncertainties in the
coefficient factors.  We find, considering only t-squark
contributions, that
\beq
{\Gamma(B^+\rightarrow \bar{K}^oK^+)\over
\Gamma(B^+\rightarrow K^+ J/\psi)}=\left({1-{m_{J/\psi}^2\over
m_B^2}}\right)^{-1}\left({f_K\over f_{J/\psi}}\right)^2
{|\ltbs\ltsd|^2(m_W/m_{\tilde{t}})^4\over (G_Fm_W^2)^2|V_{cb}|^2
|V_{cs}|^2} \times (9.8\times 10^{-2})\eeq
Here, $m_{\tilde{t}}$ is the mass of the scalar top, $m_{J/\psi}$
is the mass of the $J/\psi$ meson, $f_K$ and $f_{J/\psi}$ are the
decay constants of the $K$ and $J/\psi$, which are related to
their wave functions at the origin, and $V_{cb}$ and $V_{cs}$ are
CKM elements.  We shall use $f_{J/\psi}/f_K \simeq 2.55$. Using
the experimental branching ratio for $B^+\rightarrow K^+J/\psi$,
namely~\cite{pdg} $10.2\times 10^{-4}$, we have
\beq
{\rm B.R.}(B^+\rightarrow \bar{K}^oK^+)\simeq
1.97|\ltbs\ltsd|^2(m_W/m_{\tilde{t}})^4.\eeq

On using the recent experimental upper bound~\cite{cleo} of $5\times
10^{-5}$ on the branching ratio, and noting that one can replace
the scalar top with a scalar charm or scalar up, we then have
\beq
{|\lm_{qbs}''\lm_{qsd}''|m^2_W\over m^2_{\tilde{q}}}< 5
\times 10^{-3},\eeq
for $q=t,c,u$.  We have redone the computation of Eq. 2 using the
methods of heavy quark symmetry and find a 15\% downward revision
in the upper bound\footnote{This gives some idea of the
theoretical uncertainty in the bound.}.

One can repeat the calculation for the decay $B^+\rightarrow
\bar{K}^o\pi^+$ (or equivalently $B^-\rightarrow
{K}^o\pi^-$) in much the same way.  The result (with only the
t-squark contribution being considered) is
\beq
{\rm B.R.}(B^+\rightarrow
\bar{K}^o\pi^+)= 1.32|\ltbd\ltsd|^2(m_W/m_{\tilde{t}})^4.\eeq
Again, using the experimental upper bound~\cite{cleo} of $5\times
10^{-5}$ on the branching ratio, we have
\beq
{|\lm_{qbd}''\lm_{qsd}''|m^2_W\over m^2_{\tilde{q}}}< 4.1
\times 10^{-3},\eeq
for $q=t,c,u$.

Though these methods of calculation would be less reliable for
two-body nonleptonic decays of $D$-mesons, we can use a similar
approach there to get an order of magnitude estimate of the
corresponding bounds.  However, we find that the bounds are
significantly higher than bounds obtained from $D-\bar{D}$
mixing, discussed below.

Additional bounds were obtained by Barbieri and
Masiero~\cite{barbieri} from the contribution of $K$-$\bar{K}$
mixing to the $K_L$-$K_S$ mass difference.  There are two main box
diagrams towards this contribution, shown in Fig. 2.   Assuming
that one or the other of these diagrams is dominant, the
results of Barbieri and Masiero lead to
\footnote{We are considering bounds on the real part of the
coupling constants only.  Bounds on possible imaginary parts were
also discussed by Barbieri and Masiero.  If the couplings have
significant imaginary parts, then in the bounds in this paper arising
from $K$-$\bar{K}$ and $D$-$\bar{D}$ mixing,
 $|\lambda_a\lambda_b|$ must be replaced by $|{\rm
Re}(\lambda_a^{*2}\lambda_b^2|)|^{1/2}$.}
\beq  |\ltbs\ltbd| < {\rm min}\left(6\times
10^{-4}(m_{\tilde{t}}/m_W),
 3\times 10^{-4}(m_{\tilde{t}}/m_W)^2\right)\eeq
\beq  |\lcbs\lcbd| < {\rm min}\left(6\times
10^{-4}(m_{\tilde{c}}/m_W),
 2\times 10^{-4}(m_{\tilde{c}}/m_W)^2\right)\eeq
These authors did assume that the top quark was much lighter than
scalar top, and that all squark masses are degenerate.  The former
assumption has been invalidated by recent data, so their bounds
need to be revised.

    These results can be generalized by noting that the exchanged
quarks in Fig. 2a can be any two charge-$2/3$
quarks, and also that the exchanged quarks in Fig. 2b can
consist of one $c$-quark and one $t$-quark ($u$-quark
contributions are suppressed by a mass insertion in this diagram).
We have calculated the possible contributions, assuming that all
of the squark masses are equal except that of the scalar top,
and using a mass of $175$ GeV for the top quark with updated CKM
angles.\footnote{In the limit of small top quark mass, our
analytic result for the effective Hamiltonian is a factor of two
smaller than that given by Barbieri and Masiero.}  In Fig 2a, we
have also included (as did Barbieri and Masiero ) the contribution
arising by replacing all of the particles in the box with their
superpartners; we have not included the similar contribution from
Fig. 2b due to the extra unknown parameters (it is unlikely that
this contribution will almost exactly cancel the calculated
contributions; thus our results will not be significantly affected
by them).  We require that these contributions not exceed the
standard model contribution (which is uncertain by roughly a factor
of two), and  have plotted the upper bound for various products of
couplings in Fig. 3. It is not hard to see that the contribution
from $B$-$\bar{B}$ mixing to the
$B_L$-$B_S$ mass difference will give bounds on the same couplings,
but will be weaker.

 Bounds from $D$-$\bar{D}$ mixing can also be considered.
We find that one of the two box
diagrams is suppressed by small CKM angles and the other gives
\beq |\lcbs\lubs| < 3.1\times
10^{-3}(m_{\tilde{s}}/m_W)\eeq

What about $\Delta B = 2$ decays?   One can envision the process of
Figure 4, which will lead to the decay $B\rightarrow
\Sigma^+\Sigma^-$ or $\Lambda\Lambda$ .  However, a simple estimate
of the rate gives branching ratios (assuming the B-violating
couplings are unity and the scalar quark masses are near the W
mass) of
$O(10^{-8})$.   The smallness of the rate is due in large part
to two small CKM elements in the amplitude. Thus, such processes
will not provide interesting bounds unless
$10^{10}$ B-decays are studied.

It is generally assumed that the presence of both lepton
nonconserving terms and baryon nonconserving terms leads to
unacceptably rapid proton decay.  However, if enough third
generation fields are involved, proton decay can be sufficiently
suppressed as to make some of the bounds very weak (or, in a few
cases, nonexistent).  To see this, suppose both $\lm$ and $\lm''$
terms exist.  Consider the bound on the product
$|\lm_{\mu\tau\tau}\lusd|$.  This will lead to proton decay through
the diagram of Figure 5a.  Although there is a suppression due to
mixing angles and heavy squark propagators, the proton lifetime
bound gives a strong bound of $|\lm_{\mu\tau\tau}\lusd| < 10^{-14}$.
This bound is independent of the final state leptons, and thus
applies to all 9 of the $\lm$ couplings.  Similar bounds can be
obtained for all $\lm''$ couplings with at most one heavy field
(which is then the scalar quark); we obtain $|\lm_{ijk}\lubd| <
10^{-13}$,
$|\lm_{ijk}\lubs| < 10^{-12}$, $|\lm_{ijk}\lcsd| < 10^{-13}$ and
$|\lm_{ijk}\ltsd| < 10^{-12}$.  However, if the $\lm''$ coupling has
two heavy fields, a loop is necessary, as shown in Figure 5b.  This
gives much weaker bounds; we obtain $|\lm_{ijk}\ltbs| < 10^{-2}$,
$|\lm_{ijk}\ltbd| < 10^{-3}$,
$|\lm_{ijk}\lcbs| < 10^{-3}$ and $|\lm_{ijk}\lcbd| < 10^{-2}$.  We
thus see that the lack of obvservation of proton decay does NOT
always give very strong bounds on the product of the lepton number
violating and baryon number violating couplings.

Finally, we consider the product of $\lm'$ and $\lm''$ couplings.
Here, there are $27\times 9$ possible products, of the form
$|\lm'_{ijk}\lm''_{abc}|$.  The diagrams that can lead to proton
decay are shown in Fig. 6.  For each of these, one can have a
$\tau$ or
$c,b, t$ quark on an external leg, in which case that leg must be
virtual and decay through a $W$.  We have examined all posssible
products of couplings and found that the vast majority are tightly
bounded (product is less than
$10^{-6}$), but some are not.  Rather than list the bounds for all
243 combinations, only the bounds which are greater than $10^{-6}$
(for the product of the $\lm'$ and $\lm''$ couplings) will be
given explicitly.  It is found that all products with a $\lusd$,
$\lubd$ and $\lubs$ are smaller than $10^{-9}$.  The same is true
for $\lcsd$, $\lcbd$ and $\lcbs$, {\it except} for
$|\lm'_{tbl}\lcsd|$ which is $ < 10^{-1}$, $|\lm'_{usl}\lcbd|$
which is  $ < 10^{-2}$ and
$|\lm'_{udl}\lcbs|$, for which no bound better than the unitarity
bound could be found.  Here ${\it l}$ is any lepton.  For $\ltsd$,
the only bound which is not very small is the combination
$|\lm'_{cbl}\ltsd|$, which is bounded only by unitarity.  For
$\ltbd$ and $\ltbs$, we find $|\lm'_{ud(e,\mu)}\ltbd| < 10^{-2}$,
$|\lm'_{us(e,\mu)}\ltbs| < 10^{-2}$,
$|\lm'_{c(d,s)(e,\mu)}\ltbd| < 10^{-3}$,
$|\lm'_{c(d,s)(e,\mu)}\ltbs| < 10^{-3}$, $|\lm'_{cd\tau}\ltbd| <
10^{-5}$, $|\lm'_{cs\tau}\ltbs| < 10^{-5}$, whereas
$|\lm'_{(u,c)d\tau}\ltbs|$ and
$|\lm'_{(u,c)s\tau}\ltbd|$ are bounded only by unitarity.  All other
bounds are quite tiny.  Thus, of the 243 combinations of couplings,
thirty have bounds greater than $10^{-6}$, and
eight are completely unconstrained by the lack of observation so
far of proton decay.

Could the unconstrained couplings be bounded by rare B decays?
One can envision a diagram similar to Fig. 1 in which the upper
vertex is a $\lambda'$ vertex; this would lead to B decay into
three quarks and a lepton, such as $B\rightarrow p\tau$.  However,
in all of the above 30 couplings, the scalar quark leaving
the $\lambda'$ vertex is different from that entering the
$\lambda''$ vertex, and thus a loop will be necessary, suppressing
the rate.  As originally noted by Thorndike and Poling\cite{ed},
therefore, the bounds from proton decay will be better than those
from $B$ decay in all cases, unless branching ratios of $10^{-8}$
or better are obtained.

It is interesting that the standard supersymmetric model can
contain some baryon number and lepton number violating coupling
constants which are of order unity, and which do not lead to
excessively fast proton decay.  Such couplings could be
measured when squarks and sleptons are discovered, since they will
lead to baryon and lepton number violating squark and slepton
decays.  These have been extensively discussed by Dreiner and
Ross and by Dimopoulos, et al.~\cite{herbi}, who analyze the impact
of such decays on phenomenology.

P.R. would like to acknowledge the hospitality of the College of
William and Mary and of the CEBAF laboratory during the course of
this work.  We would like to thank Michael Peskin  and Ed
Thorndike for useful discussions, and Frank W\"urthwein and Karen
Lingel for private communications regarding soon-to-be published
bounds from CLEO.

 \def\prd#1#2#3{{\rm Phys. ~Rev. ~}{\bf D#1} (19#2) #3}
\def\plb#1#2#3{{\rm Phys. ~Lett. ~}{\bf B#1} (19#2) #3}
\def\npb#1#2#3{{\rm Nucl. ~Phys. ~}{\bf B#1} (19#2) #3}
\def\prl#1#2#3{{\rm Phys. ~Rev. ~Lett. ~}{\bf #1} (19#2) #3}

\bibliographystyle{unsrt}

\newpage

\begin{figure}

\vglue 0.1in

\caption{The dominant diagrams contributing to
$B^+\rightarrow \bar{K}^oK^+$. The $\tilde{t}$ refers to
the scalar top quark, the $g$ to the gluon.}
\end{figure}

\begin{figure}


\vglue 0.in

\caption{Contribution of R-parity violating couplings to the
$K_L$-$K_S
$ mass difference.}
\end{figure}

\begin{figure}

\vglue 0.2in

\caption{Bounds on products of $\lambda''$ couplings, plotted vs.
the scalar bottom mass, $m_s$, arising from constraints from
$K$-$\bar{K}$ mixing. The ratio of the scalar top mass
to the scalar bottom mass is taken to be
$0.5$; all other scalar quark masses are degenerate with the
scalar bottom mass.   The bounds on
$|\ltbd\ltbs|$,
$|\ltbs\lcbd|$,
 $|\lcbd\lcbs|$  are
given by the solid, dashed, and dotted lines
respectively.  The bound on
$|\lcbs\lcbd\lubs\lubd|^{1/2}$ is
$\protect\sqrt{2}$ lower than the dotted line; the bound on
$|\lcbs\lcbd\ltbd\ltbs|$ and $|\lubd\lubs\ltbd\ltbs|$
are given by the products of the solid and dotted lines, the bound
on
$|\lubd\lubs|$ is also the dotted line; and the bound on
$|\ltbd\lcbs|$ is larger than that of
$|\ltbs\lcbd|$ by a factor of roughly $1.7$. If the scalar top mass
is taken to be $0.9m_s$, then only the $|\ltbd\ltbs|$ curve
changes, increasing by roughly 50\% over the entire range. }

\end{figure}

\begin{figure}

\vglue 0.in

\caption{A typical contribution of R-parity violating
couplings to the baryon number violating decay of the $B^o$, in
this case yielding $B^o\rightarrow\Lambda\Lambda$.}
\end{figure}

\begin{figure}

\vglue 0.in

\caption{Diagram (a) shows a typical proton decay arising from
the existence of both $\lambda$ and $\lambda''$ R-parity violating
terms. In diagram (b), the contribution is given for the case in
which the $\lambda''$ term contains two heavy quarks.}
\end{figure}

\begin{figure}

\vglue 0.in

\caption{Contributions to proton decay arising from the
existence of both $\lambda'$ and $\lambda''$ R-parity violating
terms.   The tree-level diagram contributes if the $\lambda'$ and
$\lambda''$ terms each have a single, identical heavy field; in
other cases, either the loop diagrams contribute or one of the
external lines is a heavy quark which must then emit a virtual $W$
(or, in some cases, both must occur).}
\end{figure}

\end{document}